\documentstyle[prl,aps,multicol]{revtex}

\renewcommand{\narrowtext}{\begin{multicols}{2}
\global\columnwidth20.5pc} 
\renewcommand{\widetext}{\end{multicols}
\global\columnwidth42.5pc} \multicolsep = 8pt plus 4pt minus 3pt

\begin{document}
\draft
\title{Tunneling Assisted Acoustic Plasmon-Quasiparticle Excitation Resonances in
Coupled Q1D Electron Gases}
\author{Guo-Qiang Hai and Marcos R. S. Tavares}
\address{Instituto de F\'{\i}sica de S\~{a}o Carlos, Universidade de S\~{a}o Paulo,\\
13560-970 S\~{a}o Carlos, SP, Brazil}
\maketitle

\begin{abstract}
We show that a weak non-resonant tunneling between two quantum wires leads
to splitting of the acoustic plasmon mode at finite wavevector. Two gaps
open up in the dipersion of the acoustic plasmon mode when its frequency is
close to the frequencies of the quasiparticle excitations. In contrast to
the Laudau damping of the collective excitations, these gaps are attributed
to tunneling assisted acoustic plasmon$-$quasiparticle excitation
resonances. We predict that such a resonance can be observed in inelastic
light scattering spectrum.
\end{abstract}

\pacs{71.45.-d; 73.20.Mf; 73.40.Gk}

\narrowtext

The plasmons of coupled low-dimensional electron gas systems provide a
valuable platform to study the electronic many-body effects. In coupled
double one-dimensional (1D) electron quantum wires, similarly to coupled
two-dimensional electron systems\cite{1}, optical and acoustic plasmon modes 
\cite{2,3,4} were found. They were interpreted, respectively, as\ in-phase
and out-of-phase oscillations of\ the electron charge density in the two
wires. Theoretical studies\cite{3,4,5,6,7,8,9,10} have been done on the
plasmon dispersions, electron-electron correlation, far-infrared absorption,
Coulomb drag, and tunneling effects in these systems. Correlation induced
instability\cite{8,9} of the collective modes were predicted in coupled low
density quantum wires. Experimentally, far-infrared spectroscopy and Raman
scattering were used to detect the collective excitations.\cite{2,11} Very
recently, it was shown that a weak resonant tunneling in the coupled two 1D
electron gases leads to a plasmon gap in the acoustic mode at zero
wavevector.\cite{3}

In this Letter, we report a theoretical study of the effects of weak
tunneling on the collective excitations in coupled quasi-1D electron gases.
Tunneling between quantum wires can modify the collective behavior of the
electron systems in several aspects. Interwire charge transfer and
intersubband scattering become possible through the tunneling. As a
consequence, new plasmon modes and coupling between different modes appear.
On the other hand, intersubband interaction leads to intersubband
quasiparticle excitations. We expect the tunneling will mainly affect the
acoustic plasmon mode because its polarization field is localized in the
space between the two wires where the tunneling occurs. Our numerical
results of paramount importance show that a weak non-resonant tunneling
between the wires produces two gaps in the acoustic plasmon mode at finite
wavevector $q$. These gaps are attributed to the{\it \ }{\em tunneling
assisted acoustic plasmon}$-${\em quasiparticle excitation resonances}. It
means that, in contrast to the Landau damping of plasmon modes, a resonant
scattering occurs between the collective plasmon excitation and the{\em \
intersubband quasiparticle excitation} through tunneling. Such a resonance
leads to splitting of the acoustic plasmon mode around the quasiparticle
excitation region and, consequently, a double peak structure in the
corresponding inelastic light scattering spectrum.

We consider a two-dimensional system in the $xy$ plane subjected to an
additional confinement in the $y$-direction which forms two quantum wires
parallel to each other in the $x$-direction. The confinement potential in
the $y$-direction is taken to be of square well type of height $V_{b}$ and
widths $W_{1}$ and $W_{2}$ representing the first and the second wire,
respectively. The potential barrier between the two wires is of width$\
W_{b} $. The subband energies $E_{n}$ and the wave functions $\phi _{n}(y)$
are obtained from the numerical solution of the one-dimensional
Schr\"{o}dinger equation in the $y$-direction. We restrict ourselves to the
case where $n=1,2 $ and define $\omega _{0}=E_{2}-E_{1}$ as being the gap
between the two subbands. The interpretation of the index $n$\thinspace
depends on tunneling between the two wires. When there is no tunneling, $n$
is wire index. On the opposite, when the wires are in resonant tunneling
condition, $n$ is subband index.

The dispersions of the plasmon modes are obtained by the poles of the
density-density correlation function, or equivalently by the zeros of the
determinant of the dielectric matrix $\det |\varepsilon (\omega ,q)|=0$\
within the random-phase approximation (RPA). The RPA has been proved a
successful approximation in studying the collective charge excitations of
Q1D electron gas by virtue of the vanishing of all vertex corrections to the
1D irreducible polarizability.\cite{3} Fig. 1 shows the plasmon dispersions
of the coupled GaAs/Al$_{0.3}$Ga$_{0.7}$As ($V_{b}=228$ meV) quantum wires
in (a) resonant tunneling and (b) non-resonant tunneling. The numerical
results, with tunneling effects, of the in-phase (optical) $\omega _{+}$ and
out-of-phase (acoustic) $\omega _{-}$ modes are presented by the thin-solid
and thick-solid curves, respectively. \ For a comparison, the in-phase
(out-of-phase) plasmon modes without tunneling are plotted\ in the
thin-dashed (thick-dashed) curves. \ In Fig. 1(a), we observe that, in
resonant tunneling, the out-of-phase mode losses its acoustic characteristic
at small $q$ replaced by two intersubband modes. In Fig. 1(b), for the two
wires out of resonant tunneling, we find that 99.4\% of the electrons in the
lowest (second) subband are localized in the wide (narrow) quantum wire. In
other words, each quantum wire of the 1D electron gas only has a small edge
in the other. However, such an edge affects significantly the acoustic
plasmon mode. Two gaps open up around the intersubband quasiparticle
excitation region.

The dynamical dielectric function is given by $\varepsilon _{nn^{\prime
},mm^{\prime }}(\omega ,q)=\delta _{nm}\delta _{n^{\prime }m^{\prime
}}-V_{nn^{\prime },mm^{\prime }}(q)\Pi _{nn^{\prime }}(q,\omega ),$ where $%
\delta _{nm}$ is the Kronecker $\delta $ function, $V_{nn^{\prime
},mm^{\prime }}(q)$ the bare electron-electron Coulomb interaction
potential, and $\Pi _{nn^{\prime }}(\omega ,q)$ the 1D polarizability.\cite
{3,12} Within the RPA, $\Pi _{nn^{\prime }}(\omega ,q)$ is taken as the\
non-interacting irreducible polarizability function for a clean system free
from any impurity scattering. In the presence of impurity scattering, we use
Mermin's formula\cite{13} including the effect of level broadening through a
phenomenological damping constant $\gamma $. The electron-electron
interaction potential $V_{nn^{\prime },mm^{\prime }}(q)$ describes
two-particle scattering events.\cite{12,14} There are different scattering
processes in the coupled quantum wires: (i) Intrawire (intrasubband)
interactions $V_{11,11}(q)=V_{A},$ $V_{22,22}(q)=V_{B},$ and $%
V_{11,22}(q)=V_{22,11}(q)=V_{C}$ representing\ the scattering in which the
electrons keep in their original wires (subbands); (ii) Interwire
(intersubband) interactions $%
V_{12,12}(q)=V_{21,21}(q)=V_{12,21}(q)=V_{21,12}(q)=V_{D}$ representing the
scattering in which both electrons change their wire (subband) indices; and
(iii) Intra-interwire (subband) interactions$\
V_{11,12}(q)=V_{11,21}(q)=V_{12,11}(q)=V_{21,11}(q)=V_{J}$ and $%
V_{22,12}(q)=V_{22,21}(q)=V_{12,22}(q)=V_{21,22}(q)=$ $V_{H}$ indicating the
scattering in which only one of the electrons suffers the interwire
(intersubband) transition.\ Notice that, when there is no tunneling, $%
V_{D}=V_{H}=V_{J}=0$. Clearly, they are responsible for tunneling effects on
the collective excitations.

When the tunneling is considered, the plasmon dispersions of two coupled
quantum wires are determined by the equation,
\[
F_{1}F_{2}-\left[ (1-V_{A}\Pi _{11})V_{H}^{2}\Pi _{22}+\left( 1-V_{B}\Pi
_{22}\right) V_{J}^{2}\Pi _{11}\right. -
\]
\begin{equation}
\left. 2V_{C}V_{J}V_{H}\Pi _{11}\Pi _{22}\left( \Pi _{12}+\Pi _{21}\right) 
\right] =0,  \label{eq1}
\end{equation}
where $F_{1}=\left( 1-V_{A}\Pi _{11}\right) \left( 1-V_{B}\Pi _{22}\right)
-V_{C}^{2}\Pi _{11}\Pi _{22}$ and $F_{2}=\ 1-V_{D}\left( \Pi _{12}+\Pi
_{21}\right) .$\ This equation consists of two terms: $F_{1}F_{2}$ and the
rest. We know that tunneling introduces the Coulomb scattering potential $%
V_{D},$ $V_{J}$ and $V_{H}$. \ However, for two symmetric quantum wires in
resonant tunneling, $V_{J}$ and $V_{H}$ vanish and, consequently, the second
term in Eq. (1) is zero. So, the plasmon modes are determined by equations $%
F_{1}=0$ and $F_{2}=0$. The latter carries the information of tunneling
effects resulting in two out-of-phase (intersubband) modes as shown in Fig.
1(a). To reveal the relative importance of the different plasmon modes, we
performed a numerical calculation of the oscillator strength defined by $\pi
\{|\partial (\det |\varepsilon |)/\partial \omega |_{\omega =\omega _{\pm
}}\}^{-1}$. It was found that the higher frequency out-of-phase plasmon mode
is of finite oscillator strength at $q=0.$ But the lower one has a very
small oscillator strength and is unimportant.\cite{14}

When the two wires are out of resonant tunneling, the out-of-phase plasmon
mode changes dramatically at small $q$ as shown in Fig. 1(b). It restores
the acoustic behavior at $q\rightarrow 0$ but develops two gaps at finite $%
q. $\ The splitting of the acoustic plasmon mode occurs when its frequency
is close to the frequencies of the \ {\it intersubband} quasiparticle
excitations QPE$_{12}$. In this case, the small overlap between the
wavefunctions of the two subbands leads to $V_{A}$, $V_{B}$, and $V_{C}\gg
V_{D},V_{J}$ and $V_{H}.$ It means that the $F_{1}$ in Eq.(1) is now
responsible for the main features of both the optical and acoustic plasmon
modes. A numerical test indicates that the roots of the equation $F_{1}=0$
can almost recover the optical and acoustic plasmon dispersions of which
tunneling is not considered. Whereas the part$\ F_{2}$\ relating to possible
intersubband plasmon becomes less important. We also notice that $V_{D}$
does not appear in the coupling term in Eq.(1). So, the potentials $V_{J}$
and $V_{H}$ are responsible for the splitting of the acoustic plasmon mode.
These interactions represent the electron-electron scattering during which
only one of them experiences intersubband transition. When the momentum and
energy transfer between the two electrons occur in the region QPE$_{12}$,
only this electron creates an intersubband electron-hole pair. From this
point of view,\ the momentum and energy conservation in the scattering leads
to such a transition getting rid of the Landau damping. In other words, the
intra-intersubband scattering $V_{J}$ and $V_{H}$ produce a resonance
between the collective excitation and the quasiparticle excitation. From
another point of view, the scattering $V_{J}$ and $V_{H}$ result in a net
charge transfer between the wires. Thus, they produce a local electric field
between the two wires and disturb the polarization field of the acoustic
plasmon mode. The energy gaps in the acoustic plasmon mode are dependent on
the electron density and tunneling strength. We can define the gap as the
frequency difference between the lower and upper branch of the split mode at
the $q$ where the unperturbed acoustic plasmon frequency is in the center of
the quasiparticle excitation region. In Fig. 2, we show the electron density
dependence of the two gaps normalized by $\omega _{0}$ in different
structures. The energies of the two gaps decrease with increasing the total
electron density. One also sees that, for smaller barrier width, the plasmon
gaps become larger.

The plasmon modes in the coupled quantum wires can be observed in the Raman
spectroscopy. The intensity of the Raman scattering is proportional to the
imaginary part of the screened density-density correlation function with a
weight reflecting the coupling between the light and different plasmon
modes. \cite{14,15} Fig. 3 shows the calculated Raman spectra due to the
plasmon scattering of the corresponding modes in Fig. 1(b) around (a) the
lower and (b)\ the higher energy gap. In the calculation, we took the
damping constant $\gamma =0.05$ meV corresponding to a sample with electron
mobility in order of $5\times 10^{5}$ cm$^{2}$/Vs. We see a strong Raman
scattering peak at high frequencies due to the optical plasmons. Besides,
there are two split peaks due to the acoustic plasmons. With increasing $q,$
the spectral weight transfers from the lower to the higher frequency one.

Finally, we show the effects of the weak non-resonant tunneling on the
inelastic Coulomb scattering rate $\sigma _{n}(k)$ of an injected electron
in the wire $n$ with momentum $k$. The inelastic Coulomb scattering rate was
obtained by the imaginary part of the electron self-energy within the GW
approximation.\cite{14,16} In Fig. 4, we plot $\sigma _{n}(k)$\ of an
electron in the narrower quantum wire ($n=2$) of the coupled wire system
corresponding to Fig. 1(b). When the tunneling is not included, the lower
and higher scattering peaks are resulted from the emission of the acoustic
and optical plasmons, respectively. The weak tunneling influences its $k$%
-dependent behavior and leads to a splitting of the\ lower scattering peak
in $\sigma _{2}(k)$, corresponding to the splitting of the acoustic plasmon
mode.

In summary, we have studied the effects of weak tunneling on the collective
excitations in two coupled quantum wires. We show that a weak non-resonant
tunneling between the wires leads to the splitting of the acoustic plasmon
mode. Two gaps open up in the dispersion of the acoustic plasmon mode. In
contrast to the Landau damping mechanism of the collective excitations, our
result gives an evidence that the resonant coupling between the collective
excitations and the quasiparticle excitations occurs in coupled quantum
wires through tunneling. Furthermore, we predict that such a resonance can
be observed in the inelastic light scattering spectrum. Besides the optical
plasmon scattering, a double peak structure appears around the quasiparticle
excitation regime due to the split acoustic plasmon modes. The splitting of
the acoustic plasmon mode also influences other electronic properties of the
system, for instance, the Coulomb inelastic-scattering rate.

This work is supported by FAPESP\ and CNPq, Brazil.

\begin{figure}[tbp]
\caption{ Plasmon dispersions in two coupled GaAs/Al$_{0.3}$Ga$_{0.7}$As ($%
V_{b}= 228$ meV) quantum wires separated by a barrier of width $W_{b}=70$
\AA\ of (a) $W_{1}=W_{2}=150$ \AA\ and (b) $W_{1}=150$ \AA\ and $W_{2}=145$
\AA. The total electron density $N_{e}=10^{6}$ cm$^{-1}$. The solid (dash)
curves present the plasmon dispersions with (without) tunneling. The thin-
and thick-curves indicate the in-phase ($\protect\omega _{+}$)\ and
out-of-phase ($\protect\omega _{-}$) plasmon modes, respectively. The shadow
area presents the quasiparticle excitation regions QPE$_{nn^{\prime}}$.}
\end{figure}

\begin{figure}[tbp]
\caption{ The normalized gap energies as a function of the total electron
density in the coupled GaAs/Al$_{0.3}$Ga$_{0.7}$As quantum wires of (a) $%
W_{1}=150$ \AA , $W_{2}=145 $ \AA\ and $W_{b}=70$ \AA\ (solid circles with $%
\protect\omega _{0}=0.94$ meV); and (b) $W_{1}=150$ \AA , $W_{2}=140$ \AA\
and $W_{b}=50$ \AA\ (solid squares with $\protect\omega _{0}=2.01$ meV). }
\end{figure}

\begin{figure}[tbp]
\caption{Raman scattering spectra in the coupled quantum wires with $%
W_{1}=150$ \AA , $W_{2}=145$ \AA\ and $W_{b}=70$ \AA\ at different $q$: (a)
from 2 to 5$\times 10^4$cm$^{-1}$ with equivalent difference $\Delta q=$0.25$%
\times 10^4$cm$^{-1}$, and (b) from 0.4 to 1.4$\times 10^5$cm$^{-1}$ with $%
\Delta q=$0.05$\times 10^5$cm$^{-1}$. $N_{e}=10^{6}$ cm $^{-1}$ and $\protect%
\gamma =0.05$ meV. The intensity in (a) is enlarged 4 times as compared to
(b). The different curves are offset for clarity. }
\end{figure}

\begin{figure}[tbp]
\caption{Inelastic Coulomb scattering rate $\protect\sigma _{n}(k)$ in the
narrower one ($n=2$) of the coupled quantum wires corresponding to Fig.
1(b). The solid and dashed curves present the results with and without
tunneling, respectively. }
\end{figure}

\widetext

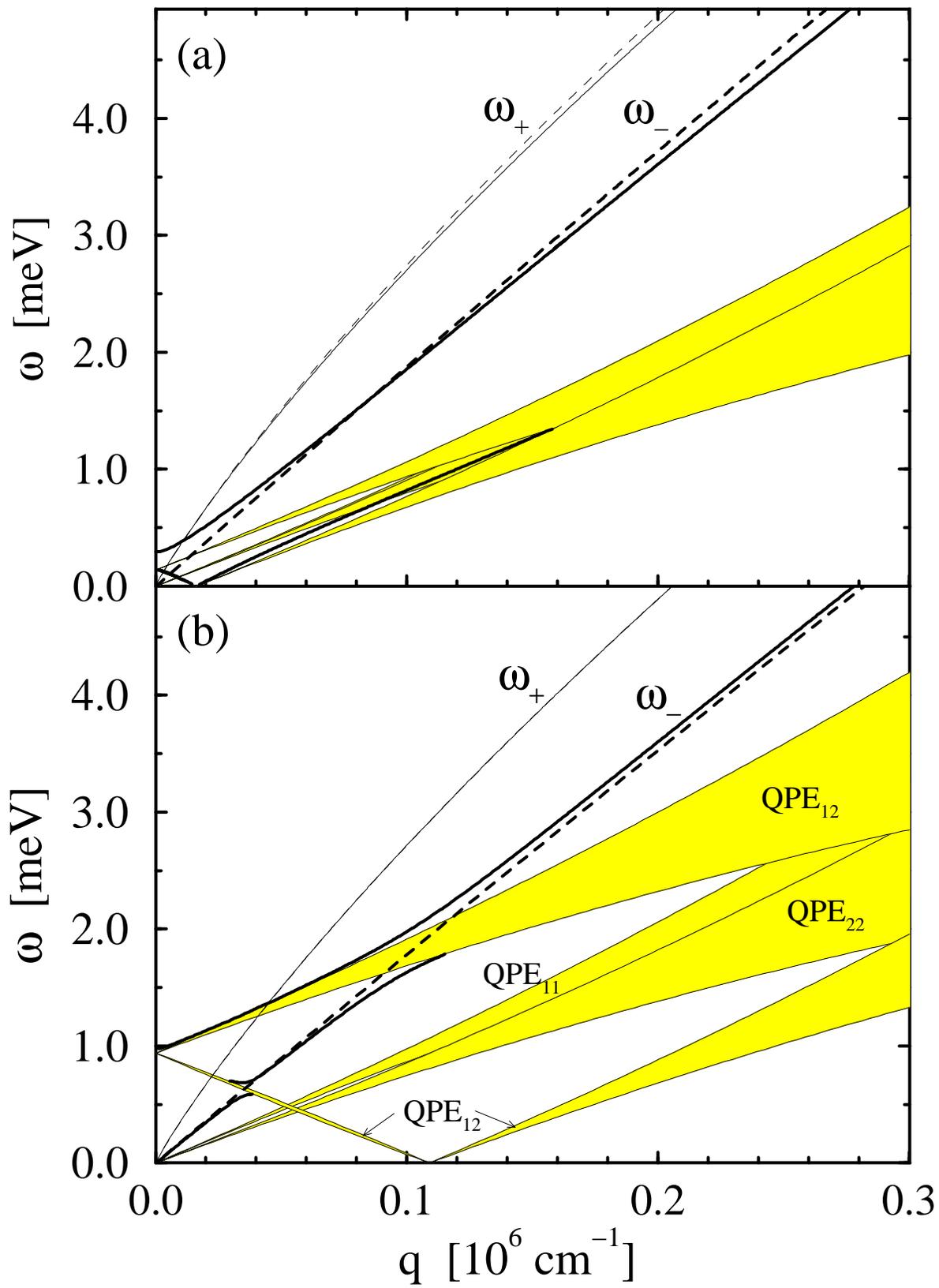
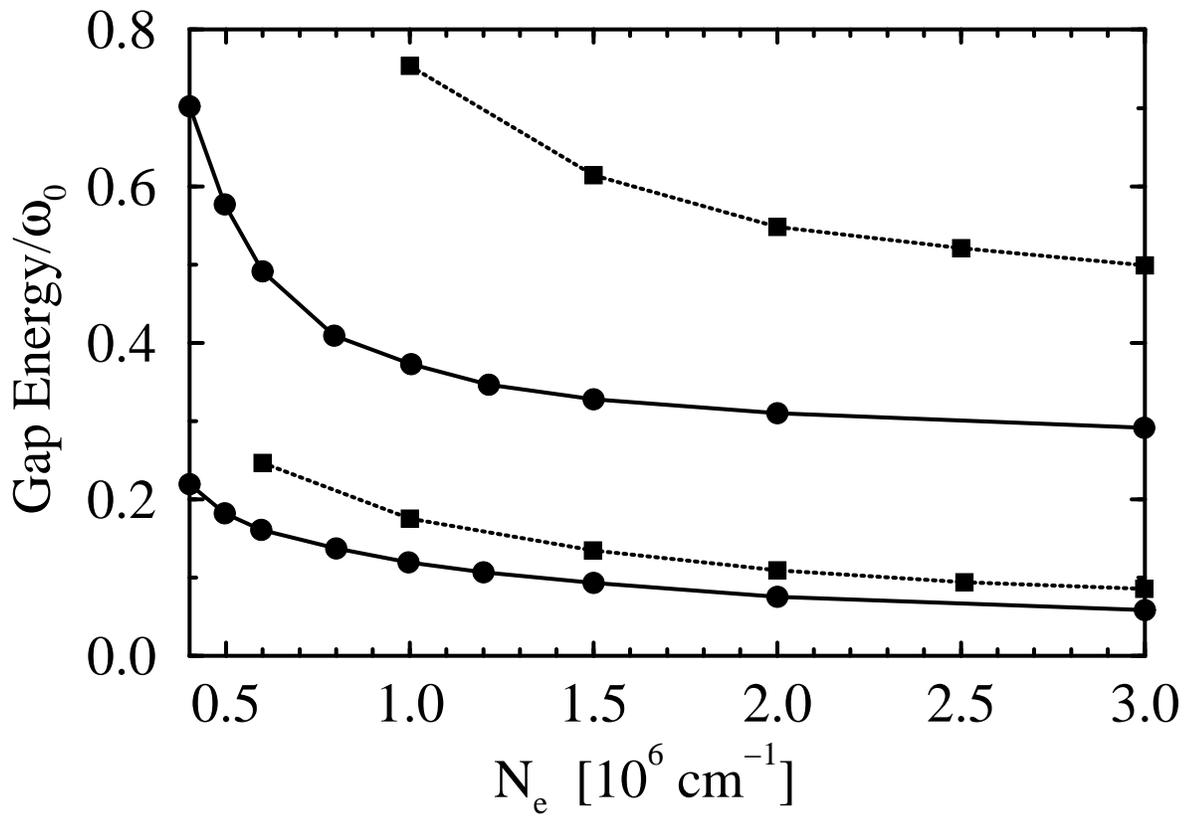
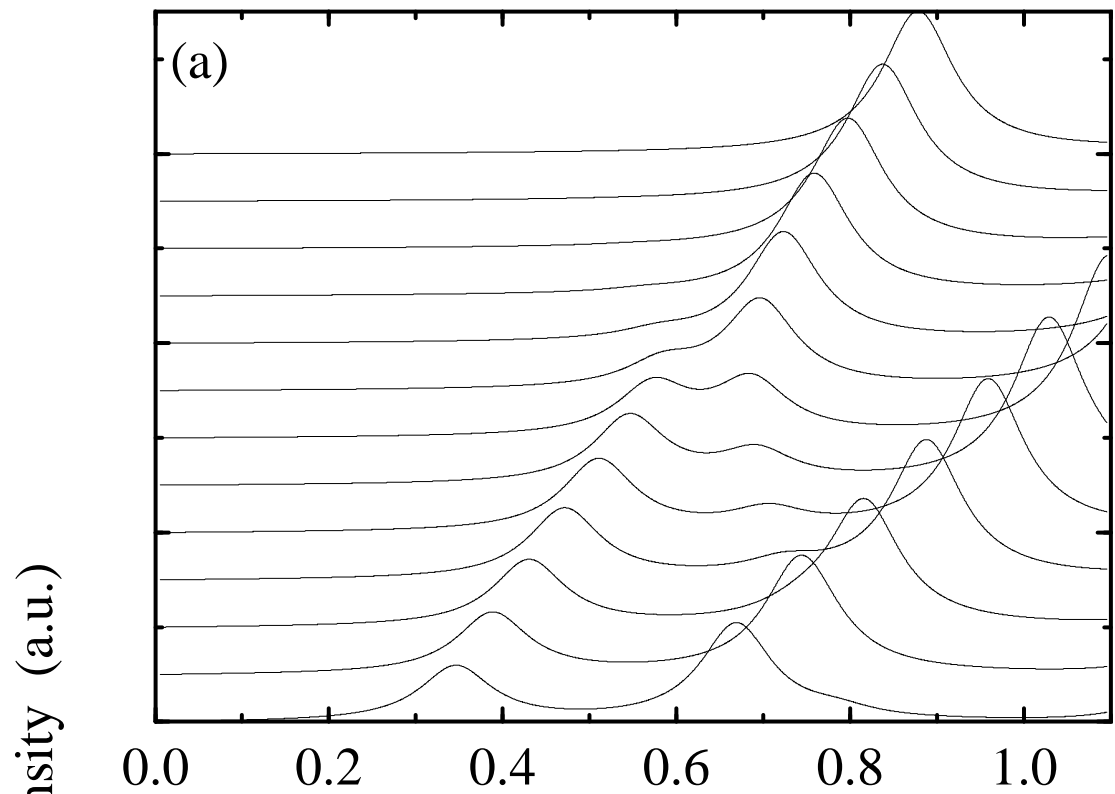
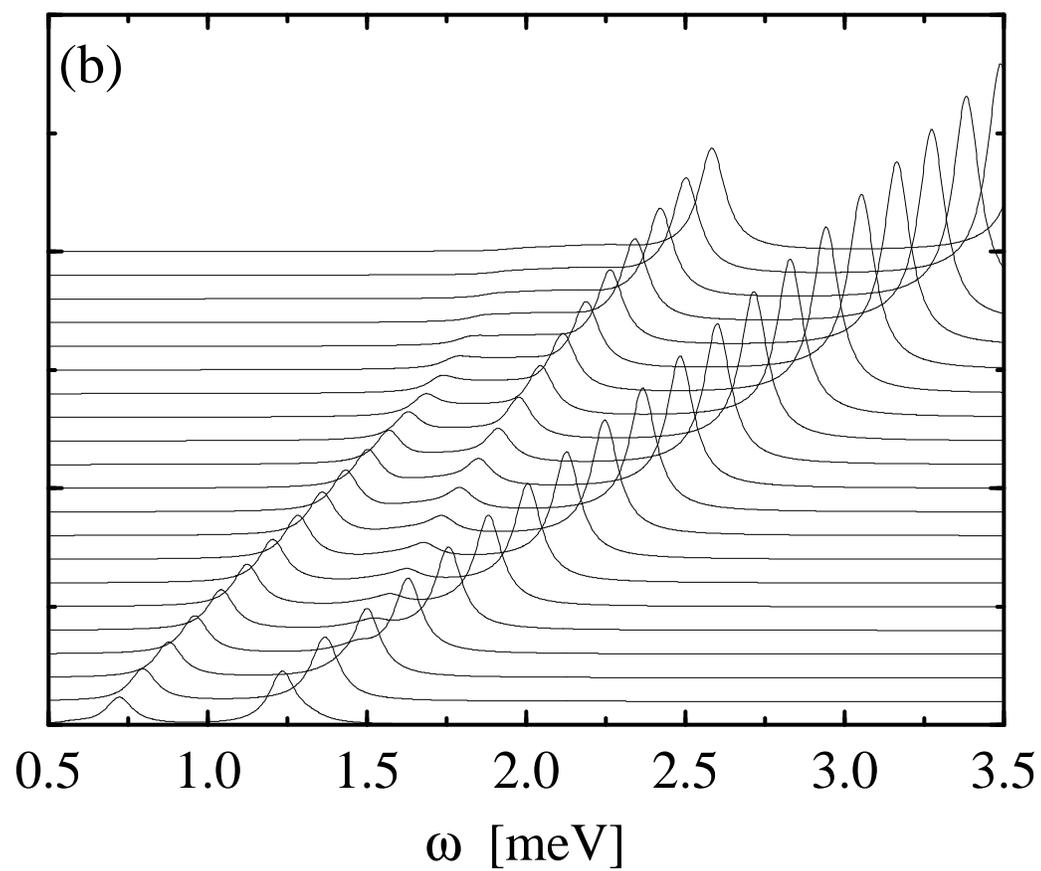
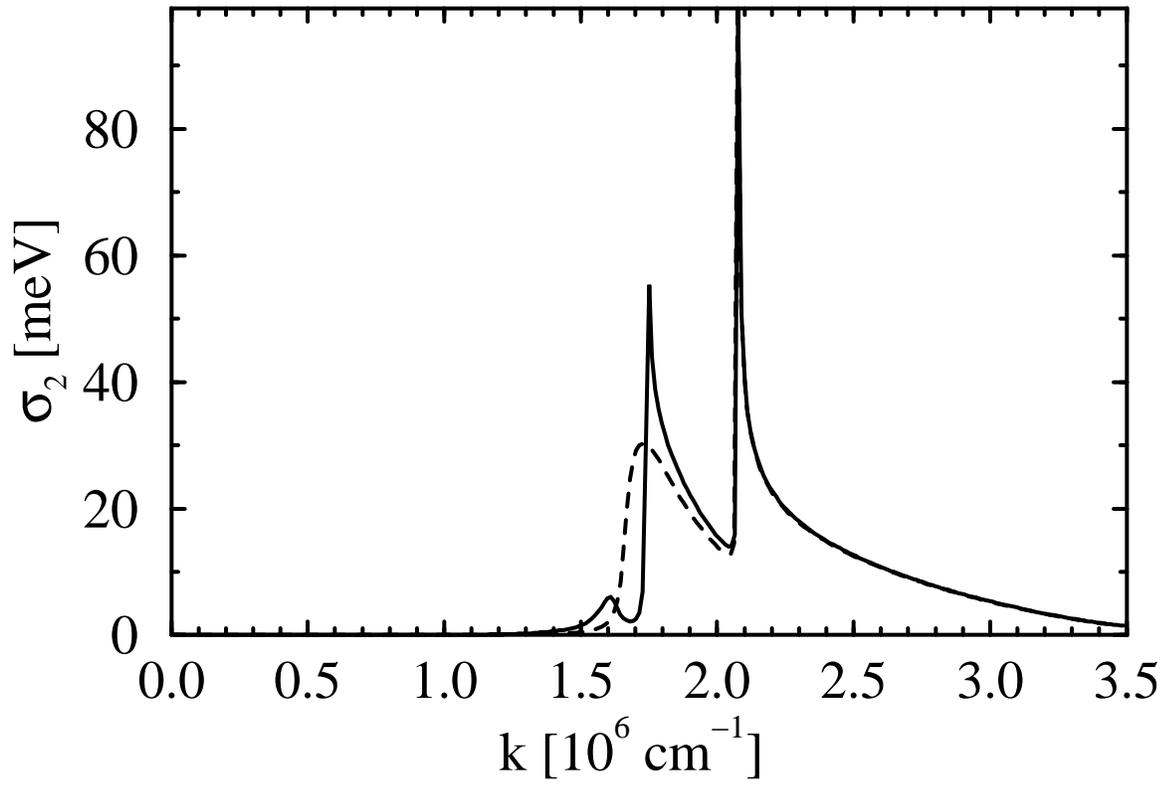
\end{document}